\documentclass[10pt,a4paper,twocolumn]{article}
\usepackage{f1000_styles}
\usepackage{hyperref}
\usepackage{enumitem}
\begin{document}

\title{Software Carpentry: Lessons Learned}
\author[1]{Greg Wilson}
\affil[1]{Mozilla Foundation / gvwilson@software-carpentry.org}

\maketitle
\thispagestyle{fancy}

\begin{abstract}

Over the last 15 years, Software Carpentry has evolved from a week-long
training course at the US national laboratories into a worldwide
volunteer effort to raise standards in scientific computing. This
article explains what we have learned along the way the challenges we
now face, and our plans for the future.

\end{abstract}
\clearpage

\section*{Introduction}

In January 2012, John Cook posted this to his widely-read blog
\cite{cook2012}:

\begin{quote}
In a review of linear programming solvers from 1987 to 2002, Bob Bixby
says that solvers benefited as much from algorithm improvements as from
Moore's law: ``Three orders of magnitude in machine speed and three
orders of magnitude in algorithmic speed add up to six orders of
magnitude in solving power. A model that might have taken a year to
solve 10 years ago can now solve in less than 30 seconds.''
\end{quote}

A million-fold speedup is impressive, but hardware and algorithms are
only two sides of the iron triangle of programming. The third is
programming itself, and while improvements to languages, tools, and
practices have undoubtedly made software developers more productive
since 1987, the speedup is percentages rather than orders of magnitude.
Setting aside the minority who do high-performance computing (HPC), the
time it takes the ``desktop majority'' of scientists to produce a new
computational result is increasingly dominated by how long it takes to
write, test, debug, install, and maintain software.

The problem is, most scientists are never taught how to do this. While
their undergraduate programs may include a generic introduction to
programming or a statistics or numerical methods course (in which
they're often expected to pick up programming on their own), they are
almost never told that version control exists, and rarely if ever shown
how to design a maintainable program in a systematic way, or how to turn
the last twenty commands they typed into a re-usable script. As a
result, they routinely spend hours doing things that could be done in
minutes, or don't do things at all because they don't know where to
start \cite{hannay2009,prabhu2011}.

This is where Software Carpentry comes in. We ran 91 workshops for
over 4300 scientists in 2013. In them, more than 100 volunteer
instructors helped attendees learn about program design, task
automation, version control, testing, and other unglamorous but
time-tested skills \cite{wilson2013}. Two independent assessments in
2012 showed that attendees are actually learning and applying at least
some of what we taught; quoting \cite{aranda2012}:

\begin{quote}
The program increases participants' computational understanding, as
measured by more than a two-fold (130\%) improvement in test scores
after the workshop. The program also enhances their habits and routines,
and leads them to adopt tools and techniques that are considered
standard practice in the software industry. As a result, participants
express extremely high levels of satisfaction with their involvement in
Software Carpentry (85\% learned what they hoped to learn; 95\% would
recommend the workshop to others).
\end{quote}

Despite these generally positive results, many researchers still find it
hard to apply what we teach to their own work, and several of our
experiments---most notably our attempts to teach online---have been
failures.

\section*{From Red to Green}

Some historical context will help explain where and why we have
succeeded and failed.

\subsection*{Version 1: Red Light}

In 1995-96, the author organized a series of articles in \emph{IEEE
Computational Science \& Engineering} titled, ``What Should Computer
Scientists Teach to Physical Scientists and Engineers?'' \cite{wilson1996}.
The articles grew out of the frustration he had working with scientists
who wanted to run before they could walk---i.e., to parallelize complex
programs that weren't broken down into self-contained functions, that
didn't have any automated tests, and that weren't under version control
\cite{wilson2006a}.

In response, John Reynders (then director of the Advanced Computing
Laboratory at Los Alamos National Laboratory) invited the author and
Brent Gorda (now at Intel) to teach a week-long course on these topics
to LANL staff. The course ran for the first time in July 1998, and was
repeated nine times over the next four years. It eventually wound down
as the principals moved on to other projects, but two valuable lessons
were learned:

\begin{enumerate}
\item
  Intensive week-long courses are easy to schedule (particularly if
  instructors are travelling) but by the last two days, attendees'
  brains are full and learning drops off significantly.
\item
  Textbook software engineering is not the right thing to teach most
  scientists. In particular, careful documentation of requirements and
  lots of up-front design aren't appropriate for people who (almost by
  definition) don't yet know what they're trying to do. Agile
  development methods, which rose to prominence during this period, are
  a less bad fit to researchers' needs, but even they are not well
  suited to the ``solo grad student'' model of working so common in
  science.
\end{enumerate}

\subsection*{Versions 2 and 3: Another Red Light}

The Software Carpentry course materials were updated and released in
2004-05 under a Creative Commons license thanks to support from the
Python Software Foundation \cite{wilson2006b}. They were used twice in
a conventional term-long graduate course at the University of Toronto
aimed at a mix of students from Computer Science and the physical and
life sciences.

The materials attracted 1000-2000 unique visitors a month, with
occasional spikes correlated to courses and mentions in other
sites. But while grad students (and the occasional faculty member)
found the course at Toronto useful, it never found an institutional
home.  Most Computer Science faculty believe this basic material is
too easy to deserve a graduate credit (even though a significant
minority of their students, particularly those coming from non-CS
backgrounds, have no more experience of practical software development
than the average physicist). However, other departments believe that
courses like this ought to be offered by Computer Science, in the same
way that Mathematics and Statistics departments routinely offer
service courses.  In the absence of an institutional mechanism to
offer credit courses at some inter-departmental level, this course,
like many other interdisciplinary courses, fell between two stools.

\begin{quote}
\textbf{It Works Too Well to be Interesting}

We have also found that what we teach simply isn't interesting to most
computer scientists. They are interested in doing research to advance
our understanding of the science of computing; things like command-line
history, tab completion, and ``select * from table'' have been around
too long, and work too well, to be publishable any longer. As long as
universities reward research first, and supply teaching last, it is
simply not in most computer scientists own best interests to offer this
kind of course.
\end{quote}

Secondly, despite repeated invitations, other people did not
contribute updates or new material beyond an occasional bug report.
Piecemeal improvement may be normal in open source development, but
Wikipedia aside, it is still rare in other fields. In particular,
people often use one another's slide decks as starting points for
their own courses, but rarely offer their changes back to the original
author in order to improve it. This is partly because educators'
preferred file formats (Word, PowerPoint, and PDF) can't be handled
gracefully by existing version control systems, but more importantly,
there simply isn't a ``culture of contribution'' in education for
projects like Software Carpentry to build on.

The most important lesson learned in this period was that while many
faculty in science, engineering, and medicine agree that their
students should learn more about computing, they \emph{won't} agree on
what to take out of the current curriculum to make room for it. A
typical undergraduate science degree has roughly 1800 hours of class
and laboratory time; anyone who wants to add more programming,
statistics, writing, or anything else must either lengthen the program
(which is financially and institutionally infeasible) or take
something out. However, everything in the program is there because it
has a passionate defender who thinks it's vitally important, and who
is likely senior to those faculty advocating the change.

\begin{quote}
\textbf{It Adds Up}

Saying, ``We'll just add a little computing to every other course,'' is
a cheat: five minutes per hour equals four entire courses in a four-year
program, which is unlikely to ever be implemented. Pushing computing
down to the high school level is also a non-starter, since that
curriculum is also full.
\end{quote}

The sweet spot for this kind of training is therefore the first two or
three years of graduate school. At that point, students have time (at
least, more time than they'll have once they're faculty) and real
problems of their own that they want to solve.

\subsection*{Version 4: Orange Light}

The author rebooted Software Carpentry in May 2010 with support from
Indiana University, Michigan State University, Microsoft, MITACS, Queen
Mary University of London, Scimatic, SciNet, SHARCNet, and the UK Met
Office. More than 120 short video lessons were recorded during the
subsequent 12 months, and six more week-long classes were run for the
backers. We also offered an online class three times (a MOOC \emph{avant
la lettre}).

This was our most successful version to date, in part because the
scientific landscape itself had changed. Open access publishing, crowd
sourcing, and dozens of other innovations had convinced scientists
that knowing how to program was now as important to doing science as
knowing how to do statistics. Despite this, though, most still
regarded it as a tax they had to pay in order to get their science
done. Those of us who teach programming may find it interesting in its
own right, but as one course participant said, ``If I wanted to be a
programmer instead of a chemist, I would have chosen computer science
as my major instead of chemistry.''

Despite this round's overall success, there were several
disappointments:

\begin{enumerate}
\item
  Once again, we discovered that five eight-hour days are more wearying
  than enlightening.
\item
  And once again, only a handful of other people contributed material,
  not least because creating videos is significantly more challenging
  than creating slides. Editing or modifying them is harder still:
  while a typo in a slide can be fixed by opening PowerPoint, making
  the change, saving, and re-exporting the PDF, inserting new slides
  into a video and updating the soundtrack seems to take at least half
  an hour regardless of how small the change is.
\item
  Most importantly, the MOOC format didn't work: only 5-10\% of those
  who started with us finished, and the majority were people who already
  knew most of the material. Both figures are in line with completion
  rates and learner demographics for other MOOCs \cite{jordan2013}, but
  are no less disappointing because of that.
\end{enumerate}

The biggest take-away from this round was the need come up with a
scalable, sustainable model. One instructor simply can't reach enough
people, and cobbling together funding from half a dozen different
sources every twelve to eighteen months is a high-risk approach.

\subsection*{Version 5: Green Light}

Software Carpentry restarted once again in January 2012 with a new
grant from the Sloan Foundation, and backing from the Mozilla
Foundation. This time, the model was two-day intensive workshops like
those pioneered by The Hacker Within, a grassroots group of grad
students helping grad students at the University of Wisconsin --
Madison.

Shortening the workshops made it possible for more people to attend,
and increased the proportion of material they retained. It also forced
us to think much harder about what skills scientists really
needed. Out went object-oriented programming, XML, Make, GUI
construction, design patterns, and software development
lifecycles. Instead, we focused on a handful of tools (discussed in
the next section) that let us introduce higher-level concepts without
learners really noticing.

Reaching more people also allowed us to recruit more instructors from
workshop participants, which was essential for scaling. Switching to a
``host site covers costs'' model was equally important: we still need
funding for the coordinator positions (the author and two part-time
administrative assistants at Mozilla, and part of one staff member's
time at the Software Sustainability Institute in the UK), but our
other costs now take care of themselves.

Our two-day workshops have been an unqualified success. Both the number
of workshops, and the number of people attending, have grown steadily:

\begin{figure}
\centering
\includegraphics[width=0.4\textwidth]{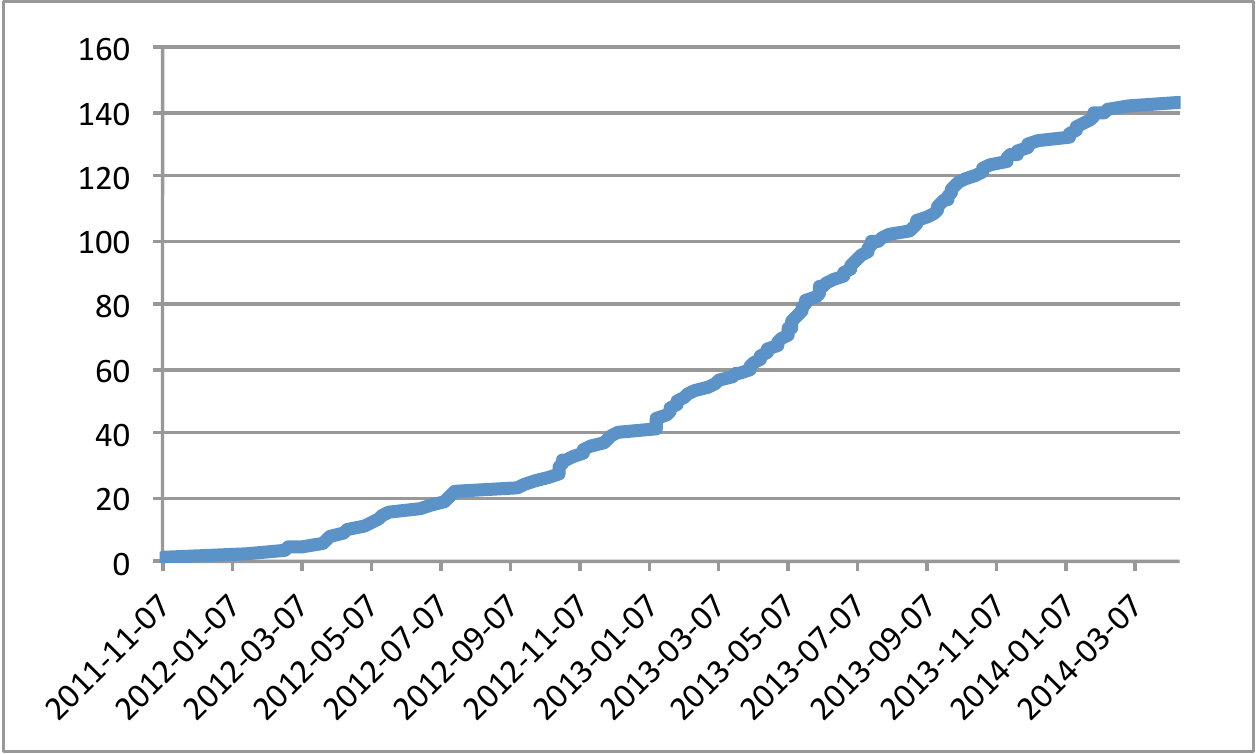}
\caption{\label{f:workshops}Cumulative Number of Workshops}
\end{figure}

\begin{figure}
\centering
\includegraphics[width=0.4\textwidth]{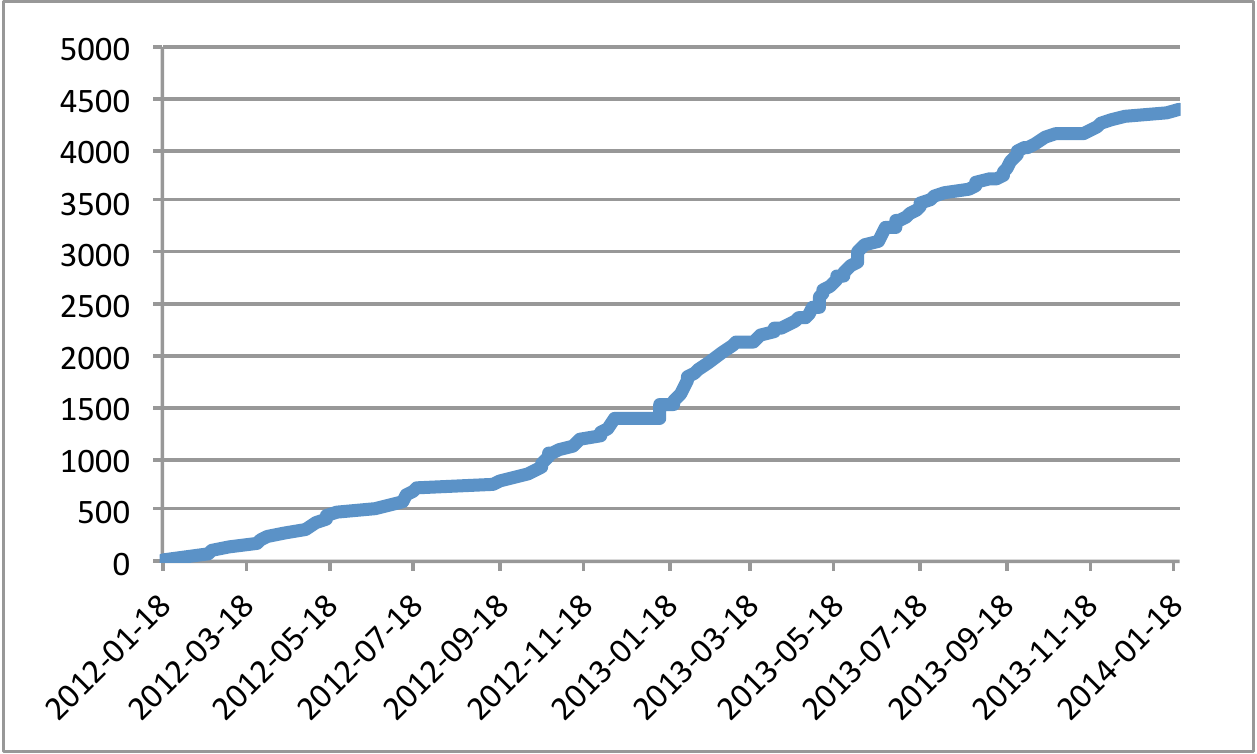}
\caption{\label{f:enrolment}Cumulative Enrolment}
\end{figure}

More importantly, feedback from participants is strongly positive. While
there are continuing problems with software setup and the speed of
instruction (discussed below), 80-90\% of attendees typically report
that they were glad they attended and would recommend the workshops to
colleagues.

\section*{What We Do}

So what does a typical workshop look like?

\begin{itemize}
\item
  \emph{Day 1 a.m.}: The Unix shell. We only show participants a dozen
  basic commands; the real aim is to introduce them to the idea of
  combining single-purpose tools (via pipes and filters) to achieve
  desired effects, and to getting the computer to repeat things (via
  command completion, history, and loops) so that people don't have
  to.
\item
  \emph{Day 1 p.m.}: Programming in Python (or sometimes R). The real
  goal is to show them when, why, and how to grow programs
  step-by-step as a set of comprehensible, reusable, and testable
  functions.
\item
  \emph{Day 2 a.m.}: Version control. We begin by emphasizing how it's
  a better way to back up files than creating directories with names
  like ``final'', ``really\_final'', ``really\_final\_revised'', and
  so on, then show them that it's also a better way to collaborate
  than FTP or Dropbox.
\item
  \emph{Day 2 p.m.}: Using databases and SQL.  The real goal is to
  show them what structured data actually is---in particular, why
  atomic values and keys are important---so that they will understand
  why it's important to store information this way.
\end{itemize}

As the comments on the bullets above suggest, our real aim isn't to
teach Python, Git, or any other specific tool: it's to teach
\emph{computational competence}. We can't do this in the abstract:
people won't show up for a hand-waving talk, and even if they do, they
won't understand. If we show them how to solve a specific problem with
a specific tool, though, we can then lead into a larger discussion of
how scientists ought to develop, use, and curate software.

We also try to show people how the pieces fit together: how to write a
Python script that fits into a Unix pipeline, how to automate unit
tests, etc. Doing this gives us a chance to reinforce ideas, and also
increases the odds of them being able to apply what they've learned
once the workshop is over.

Of course, there are a lot of local variations around the template
outlined above. Some instructors still use the command-line Python
interpreter, but a growing number have adopted the
\href{http://ipython.org/notebook.html}{IPython Notebook}, which has
proven to be an excellent teaching and learning environment.

We have also now run several workshops using R instead of Python, and
expect this number to grow. While some people feel that using R instead
of Python is like using feet and pounds instead of the metric system, it
is the \emph{lingua franca} of statistical computing, particularly in
the life sciences. A handful of workshops also cover tools such as
LaTeX, or domain-specific topics such as audio file processing. We hope
to do more of the latter going forward now that we have enough
instructors to specialize.

We aim for no more than 40 people per room at a workshop, so that
every learner can receive personal attention when needed.  Where
possible, we now run two or more rooms side by side, and use a
pre-assessment questionnaire as a sorting hat to stream learners by
prior experience, which simplifies teaching and improves their
experience.  We do \emph{not} to shuffle people from one room to
another between the first and second day: with the best
inter-instructor coordination in the world, it still results in
duplication, missed topics, and jokes that make no sense.

Our workshops were initially free, but we now often have a small
registration fee (typically \$20--40), primarily because it reduces
the no-show rate from a third to roughly 5\%.  When we do this, we
must be very careful not to trip over institutional rules about
commercial use of their space: some universities will charge us
hundreds or thousands of dollars per day for using their classrooms if
any money changes hands at any point.  We have also experimented with
refundable deposits, but the administrative overheads were
unsustainable.

\begin{quote}
\textbf{Commercial Offerings}

Our material is all covered by the Creative Commons -- Attribution
license, so anyone who wants to use it for corporate training can do
so without explicit permission from us. We encourage this: it would be
great if graduate students could help pay their bills by sharing what
they know, in the way that many programmers earn part or all of their
living from working on open source software.

What \emph{does} require permission is use of our name and logo, both
of which are trademarked. We're happy to give that permission if we've
certified the instructor and have a chance to double-check the
content, but we do want a chance to check: we have had instances of
people calling something ``Software Carpentry'' when it had nothing to
do with what we usually teach. We've worked hard to create material
that actually helps scientists, and to build some name recognition
around it, and we'd like to make sure our name continues to mean
something.
\end{quote}

As well as instructors, we rely local helpers to wander the room and
answer questions during practicals. These helpers may be participants in
previous workshops who are interested in becoming instructors, grad
students who've picked up some or all of this on their own, or members
of the local open source community; where possible, we aim to have at
least one helper for every eight learners.

We find workshops go a lot better if people come in groups (e.g., 4--5
people from one lab) or have other pre-existing ties (e.g., the same
disciplinary background). They are less inhibited about asking
questions, and can support each other (morally and technically) when
the time comes to put what they've learned into practice after the
workshop is over. Group signups also yield much higher turnout from
groups that are otherwise often under-represented, such as women and
minority students, since they know in advance that they will be in a
supportive environment.

\section*{Small Things Add Up}

As in chess, success in teaching often comes from the accumulation of
seemingly small advantages. Here are a few of the less significant
things we do that we believe have contributed to our success.

\subsection*{Live Coding}

We use live coding rather than slides: it's more convincing, it enables
instructors to be more responsive to ``what if?'' questions, and it
facilitates lateral knowledge transfer (i.e., people learn more than we
realized we were teaching them by watching us work). This does put more
of a burden on instructors than a pre-packaged slide deck, but most find
it more fun.

\subsection*{Open Everything}

Our grant proposals, mailing lists, feedback from workshops, and
everything else that isn't personally sensitive is out in the open.
While we can't prove it, we believe that the fact that people can see us
actively succeeding, failing, and learning buys us some credibility and
respect.

\subsection*{Open Lessons}

This is an important special case of the previous point. Anyone who
wants to use our lessons can take what we have, make changes, and offer
those back by sending us a pull request on GitHub. As mentioned earlier,
this workflow is still foreign to most educators, but it is allowing us
to scale and adapt more quickly and more cheaply than the centralized
approaches being taken by many high-profile online education ventures.

\subsection*{Use What We Teach}

We also make a point of eating our own cooking, e.g., we use GitHub for
our web site and to plan workshops. Again, this buys us credibility,
and gives instructors a chance to do some hands-on practice with the
things they're going to teach. The (considerable) downside is that it
can be quite difficult for newcomers to contribute material; we are
therefore working to streamline that process.

\subsection*{Meet the Learners on Their Own Ground}

Learners tell us that it's important to them to leave the workshop
with their own working environment set up. We therefore continue to
teach on all three major platforms (Linux, Mac OS X, and Windows),
even though it would be simpler to require learners to use just
one. We have experimented with virtual machines on learners' computers
to reduce installation problems, but those introduce problems of their
own: older or smaller machines simply aren't fast enough.  We have
also tried using VMs in the cloud, but this makes us dependent on
university-quality WiFi\ldots{}

\subsection*{Collaborative Note-Taking}

We often use \href{http://etherpad.org}{Etherpad} for collaborative
note-taking and to share snippets of code and small data files with
learners. (If nothing else, it saves us from having to ask students to
copy long URLs from the presenter's screen to their computers.) It is
almost always mentioned positively in post-workshop feedback, and
several workshop participants have started using it in their own
teaching.

We are still trying to come up with an equally good way to share larger
files dynamically as lessons progress.  Version control does \emph{not}
work, both because our learners are new to it (and therefore likely to
make mistakes that affect classmates) and because classroom WiFi
frequently can't handle a flurry of multi-megabyte downloads.

\subsection*{Sticky Notes and Minute Cards}

Giving each learner two sticky notes of different colors allows
instructors to do quick true/false questions as they're teaching. It
also allows real-time feedback during hands-on work: learners can put a
green sticky on their laptop when they have something done, or a red
sticky when they need help. We also use them as minute cards: before
each break, learners take a minute to write one thing they've learned on
the green sticky, and one thing they found confusing (or too fast or too
slow) on the red sticky. It only takes a couple of minutes to collate
these, and allows instructors to adjust to learners' interests and
speed.

\subsection*{Pair Programming}

Pairing is a good practice in real life, and an even better way to
teach: partners can not only help each other out during the practical,
but clarify each other's misconceptions when the solution is presented,
and discuss common research interests during breaks. To facilitate it,
we strongly prefer flat seating to banked (theater-style) seating; this
also makes it easier for helpers to reach learners who need assistance.

\subsection*{Keep Experimenting}

We are constantly trying out new ideas (though not always on purpose).
Among our current experiments are:

\begin{description}

\item[\emph{Partner and Adapt}] We have built a very fruitful
  partnership with the Software Sustainability Institute (SSI), who
  now manage our activities in the UK, and are adapting our general
  approach to meet particular local needs.

\item[\emph{A Driver's License for HPC}] As another example of this
  collaboration, we are developing a ``driver's license'' for
  researchers who wish to use the DiRAC HPC facility. During several
  rounds of beta testing, we have refined an hour-long exam to assess
  people's proficiency with the Unix shell, testing, Makefiles, and
  other skills. This exam was deployed in the fall of 2013, and we
  hope to be able to report on it by mid-2014.

\item[\emph{New Channels}] On June 24-25, 2013, we ran our first
  workshop for women in science, engineering, and medicine. This event
  attracted 120 learners, 9 instructors, a dozen helpers, and direct
  sponsorship from several companies, universities, and non-profit
  organizations. Our second such workshop will run in March 2014, and
  we are exploring ways to reach other groups that are
  underrepresented in computing.

\item[\emph{Smuggling It Into the Curriculum}] Many of our instructors
  also teach regular university courses, and several of them are now
  using part or all of our material as the first few lectures in
  them. We strongly encourage this, and would welcome a chance to work
  with anyone who wishes to explore this themselves.

\end{description}

\section*{Instructor Training}

To help people teach, we now run an
\href{http://teaching.software-carpentry.org}{online training course}
for would-be instructors. It takes 2--4 hours/week of their time for
12--14 weeks (depending on scheduling interruptions), and introduces
them to the basics of educational psychology, instructional design,
and how these things apply to teaching programming. It's necessarily
very shallow, but most participants report that they find the material
interesting as well as useful.

Why do people volunteer as instructors?

\begin{description}

\item[\emph{To make the world a better place.}]  The two things we
  need to get through the next hundred years are more science and more
  courage; by helping scientists do more in less time, we are helping
  with the former.

\item[\emph{To make their own lives better.}]  Our instructors are
  often asked by their colleagues to help with computing problems.
  The more those colleagues know, the more interesting those requests
  are.

\item[\emph{To build a reputation.}]  Showing up to run a workshop is
  a great way for people to introduce themselves to colleagues, and to
  make contact with potential collaborators. This is probably the most
  important reason from Software Carpentry's point of view, since it's
  what makes our model sustainable.

\item[\emph{To practice teaching.}]
  This is also important to people contemplating academic careers.
 
\item[\emph{To help diversify the pipeline.}]  Computing is 12-15\%
  female, and that figure has been \emph{dropping} since the
  1980s. While figures on female participation in computational
  science are hard to come by, a simple head count shows the same
  gender skew. Some of our instructors are involved in part because
  they want to help break that cycle by participating in activities
  like our workshop for women in science and engineering in Boston in
  June 2013.

\item[\emph{To learn new things, or learn old things in more detail.}]
  Working alongside an instructor with more experience is a great way
  to learn more about the tools, as well as about teaching.

\item[\emph{It's fun.}]  Our instructors get to work with smart people
  who actually want to be in the room, and don't have to mark anything
  afterward. It's a refreshing change from teaching undergraduate
  calculus\ldots{}

\end{description}

\section*{TODO}

We've learned a lot, and we're doing a much better job of reaching and
teaching people than we did eighteen months ago, but there are still
many things we need to improve.

\subsection*{Too Slow \emph{and} Too Fast}

The biggest challenge we face is the diversity of our learners'
backgrounds and skill levels. No matter what we teach, and how fast or
how slow we go, 20\% or more of the room will be lost, and there's a
good chance that a different 20\% will be bored.

The obvious solution is to split people by level, but if we ask them
how much they know about particular things, they regularly under- or
over-estimate their knowledge.  We have therefore developed a short
pre-assessment questionnaire (listed in the appendix) that asks them
whether they could accomplish specific tasks.  While far from perfect,
it seems to work well enough for our purposes.

\subsection*{Finances}

Our second-biggest problem is financial sustainability. The ``host
site covers costs'' model allows us to offer more workshops, but does
not cover the 2 full-time equivalent coordinating positions at the
center of it all.  We do ask host sites to donate toward these costs,
but are still looking for a long-term solution.

\subsection*{Long-Term Assessment}

Third, while we believe we're helping scientists, we have not yet done
the long-term follow-up needed to prove this. This is partly because of
a lack of resources, but it is also a genuinely hard problem: no one
knows how to measure the productivity of programmers, or the
productivity of scientists, and putting the two together doesn't make
the unknowns cancel out.

What we've done so far is collect verbal feedback at the end of every
workshop (mostly by asking attendees what went well and what didn't)
and administer surveys immediately before and afterwards. Neither has
been done systematically, though, which limits the insight we can
actually glean. We are taking steps to address that, but the larger
question of what impact we're having on scientists' productivity still
needs to be addressed.

\begin{quote}
\textbf{Meeting Our Own Standards}

One of the reasons we need to do long-term follow-up is to find out for
our own benefit whether we're teaching the right things the right way.
As just one example, some of us believe that Subversion is significantly
easier for novices to understand than Git because there are fewer places
data can reside and fewer steps in its normal workflow. Others believe
just as strongly that there is no difference, or that Git is actually
easier to learn. While learnability isn't the only concern---the large
social network centered around GitHub is a factor as well---we would
obviously be able to make better decisions if we had more quantitative
data to base them on.
\end{quote}

\subsection*{``Is It Supposed to Hurt This Much?''}

Fourth, getting software installed is often harder than using it. This
is a hard enough problem for experienced users, but almost by definition
our audience is \emph{inexperienced}, and our learners don't (yet) know
about system paths, environment variables, the half-dozen places
configuration files can lurk on a modern system, and so on. Combine that
with two version of Mac OS X, three of Windows, and two oddball Linux
installations, and it's almost inevitable that every time we introduce a
new tool, it won't work as expected (or at all) for at least one person
in the room. Detailed documentation has not proven effective: some
learners won't read it (despite repeated prompting), and no matter how
detailed it is, it will be incomprehensible to some, and lacking for
others.

\begin{quote}
\textbf{Edit This}

And while it may seem like a trivial thing, editing text is always
harder than we expect. We don't want to encourage people to use naive
editors like Notepad, and the two most popular legacy editors on Unix
(Vi and Emacs) are both usability nightmares. We now recommend a
collection of open and almost-open GUI editors, but it remains a
stumbling block.
\end{quote}

\subsection*{Teaching on the Web}

Challenge \#5 is to move more of our teaching and follow-up online. We
have tried several approaches, from MOOC-style online-only offerings to
webcast tutorials and one-to-one online office hours via VoIP and
desktop sharing. In all cases, turnout has been mediocre at the start
and dropped off rapidly. The fact that this is true of most high-profile
MOOCs as well is little comfort\ldots{}

\subsection*{What vs.~How}

Sixth on our list is the tension between teaching the ``what'' and the
``how'' of programming. When we teach a scripting language like Python,
we have to spend time up front on syntax, which leaves us only limited
time for the development practices that we really want to focus on, but
which are hard to grasp in the abstract. By comparison, version control
and databases are straightforward: what you see is what you do is what
you get.

We also don't as good a job as we would like teaching testing. The
mechanics of unit testing with an xUnit-style framework are
straightforward, and it's easy to come up with representative test cases
for things like reformatting data files, but what should we tell
scientists about testing the numerical parts of their applications? Once
we've covered floating-point roundoff and the need to use ``almost
equal'' instead of ``exactly equal'', our learners quite reasonably ask,
``What should I use as a tolerance for my computation?'' for which
nobody has a good answer.

\subsection*{Standardization vs.~Customization}

What we \emph{actually} teach varies more widely than the content of
most university courses with prescribed curricula. We think this is a
strength---one of the reasons we recruit instructors from among
scientists is so that they can customize content and delivery for
local needs---but we need to be more systematic about varying on
purpose rather than by accident.

\subsection*{Watching vs.~Doing}

Finally, we try to make our teaching as interactive as possible, but
we still don't give learners hands-on exercises as frequently as we
should.  We also don't give them as diverse a range of exercises as we
should, and those that we do give are often at the wrong level. This
is partly due to a lack of time, but disorganization is also a factor.

There is also a constant tension between having students do realistic
exercises drawn from actual scientific workflows, and giving them tasks
that are small and decoupled, so that failures are less likely and don't
have knock-on effects when they occur. This is exacerbated by the
diversity of learners in the typical workshop, though we hope that will
diminish as we organize and recruit along disciplinary lines instead of
geographically.

\subsection*{Better Teaching Practices}

Computing education researchers have learned a lot in the past two
decades about why people find it hard to learn how to program, and how
to teach them more effectively
\cite{guzdial2010,guzdial2013,hazzan2011,porter2013,sorva2012}.  We
do our best to cover these ideas in our instructor training program,
but are less good about actually applying them in our workshops.

\section*{Conclusions}

To paraphrase William Gibson, the future is already here---it's just
that the skills needed to implement it aren't evenly distributed. A
small number of scientists can easily build an application that scours
the web for recently-published data, launch a cloud computing node to
compare it to home-grown data sets, and push the result to a GitHub
account; others are still struggling to free their data from Excel and
figure out which of the nine backup versions of their paper is the one
they sent for publication.

The fact is, it's hard for scientists to do the cool things their
colleagues are excited about without basic computing skills, and
impossible for them to know what other new things are possible. Our
ambition is to change that: not just to make scientists more productive
today, but to allow them to be part of the changes that are transforming
science in front of our eyes. If you would like to help, we'd like to
hear from you.

\subsection*{Competing Interests}

The author is an employee of the Mozilla Foundation. Over the years,
Software Carpentry has received support from:

\begin{itemize}
\item
  The Sloan Foundation
\item
  Microsoft
\item
  NumFOCUS
\item
  Continuum Analytics
\item
  Enthought
\item
  The Python Software Foundation
\item
  Indiana University
\item
  Michigan State University
\item
  MITACS
\item
  The Mozilla Foundation
\item
  Queen Mary University London
\item
  Scimatic Inc.
\item
  SciNET
\item
  SHARCNET
\item
  The UK Met Office
\item
  The MathWorks
\item
  Los Alamos National Laboratory
\item
  Lawrence Berkeley National Laboratory
\end{itemize}

\subsection*{Grant Information}

Software Carpentry is currently supported by a grant from the Sloan
Foundation.

\subsection*{Acknowledgements}

The author wishes to thank Brent Gorda, who helped create Software
Carpentry sixteen years ago; the hundreds of people who have helped
organize and teach workshops over the years; and the thousands of
people who have taken a few days to learn how to get more science
done in less time, with less pain.  Particular thanks go to the
following for their comments, corrections, and inspiration:

\begin{itemize}
\item
  Azalee Bostroem (Space Telescope Science Institute)
\item
  Chris Cannam (Queen Mary, University of London)
\item
  Stephen Crouch (Software Sustainability Institute)
\item
  Matt Davis (Datapad, Inc.)
\item
  Luis Figueira (King's College London)
\item
  Richard ``Tommy'' Guy (Microsoft)
\item
  Edmund Hart (University of British Columbia)
\item
  Neil Chue Hong (Software Sustainability Institute)
\item
  Katy Huff (University of Wisconsin)
\item
  Michael Jackson (Edinburgh Parallel Computing Centre)
\item
  W.\ Trevor King (Drexel University)
\item
  Justin Kitzes (University of California, Berkeley)
\item
  Stephen McGough (University of Newcastle)
\item
  Lex Nederbragt (University of Oslo)
\item
  Tracy Teal (Michigan State University)
\item
  Ben Waugh (University College London)
\item
  Lynne J.\ Williams (Rotman Research Institute)
\item
  Ethan White (Utah State University)
\end{itemize}

\nocite{*}
{\small\bibliographystyle{unsrt}
\bibliography{software-carpentry-lessons-learned}}

\begin{thebibliography}{10}

\bibitem{cook2012}
John~D. Cook.
\newblock {Moore's Law Squared}, 2012.
\newblock Viewed July 2013.

\bibitem{hannay2009}
Jo~Erskine Hannay, Hans~Petter Langtangen, Carolyn MacLeod, Dietmar Pfahl,
  Janice Singer, and Greg Wilson.
\newblock How do scientists develop and use scientific software?
\newblock In {\em Second International Workshop on Software Engineering for
  Computational Science and Engineering (SECSE09)}, 2009.

\bibitem{prabhu2011}
Prakash Prabhu, Thomas~B. Jablin, Arun Raman, Yun Zhang, Jialu Huang, Hanjun
  Kim, Nick~P. Johnson, Feng Liu, Soumyadeep Ghosh, Stephen Beard, Taewook Oh,
  Matthew Zoufaly, David Walker, and David~I. August.
\newblock A survey of the practice of computational science.
\newblock In {\em Proceedings of the 24th ACM/IEEE Conference on High
  Performance Computing, Networking, Storage and Analysis}, 2011.

\bibitem{wilson2013}
Greg Wilson, D.~A. Aruliah, C.~Titus Brown, Neil P.~Chue Hong, Matt Davis,
  Richard~T. Guy, Steven~H.D. Haddock, Kathryn~D. Huff, Ian~M. Mitchell,
  Mark~D. Plumbley, Ben Waugh, Ethan~P. White, and Paul Wilson.
\newblock Best practices for scientific computing.
\newblock {\em PLoS Biology}, 12(1):e1001745, January 2014.

\bibitem{aranda2012}
Jorge Aranda.
\newblock {Software Carpentry Assessment Report}, 2012.

\bibitem{wilson1996}
Gregory~V. Wilson.
\newblock {What Should Computer Scientists Teach to Physical Scientists and
  Engineers?}
\newblock {\em IEEE Computational Science and Engineering}, Summer and Fall
  1996.

\bibitem{wilson2006a}
Greg Wilson.
\newblock {Where's the Real Bottleneck in Scientific Computing?}
\newblock {\em American Scientist}, January-February 2006.

\bibitem{wilson2006b}
Greg Wilson.
\newblock {Software Carpentry: Getting Scientists to Write Better Code by
  Making Them More Productive}.
\newblock {\em Computing in Science \& Engineering}, November-December 2006.

\bibitem{jordan2013}
Katy Jordan.
\newblock {MOOC} completion rates: The data, 2013.
\newblock Viewed July 2013.

\bibitem{guzdial2010}
Mark Guzdial.
\newblock Why is it so hard to learn to program?
\newblock In Andy Oram and Greg Wilson, editors, {\em Making Software: What
  Really Works, and Why We Believe It}, pages 111--124. O'Reilly Media, 2010.

\bibitem{guzdial2013}
Mark Guzdial.
\newblock Exploring hypotheses about media computation.
\newblock In {\em Proc.\ Ninth Annual International ACM Conference on
  International Computing Education Research}, ICER'13, pages 19--26. ACM,
  2013.

\bibitem{hazzan2011}
Orit Hazzan, Tami Lapidot, and Noa Ragonis.
\newblock {\em Guide to Teaching Computer Science: An Activity-Based Approach}.
\newblock Springer, 2011.

\bibitem{porter2013}
Leo Porter, Mark Guzdial, Charlie McDowell, and Beth Simon.
\newblock Success in introductory programming: What works?
\newblock {\em Communications of the {ACM}}, 56(8), 2013.

\bibitem{sorva2012}
Juha Sorva.
\newblock {\em Visual Program Simulation in Introductory Programming
  Education}.
\newblock PhD thesis, Aalto University, 2012.

\end{thebibliography}

\appendix

\section{Pre-Assessment Questionnaire}

\begin{itemize}

\item
  What is your career stage?
  \begin{itemize}[noitemsep]
    \item Undergraduate
    \item Graduate
    \item Post-doc
    \item Faculty
    \item Industry
    \item Support Staff
    \item Other:
  \end{itemize}

\item
  What is your discipline?
  \begin{itemize}[noitemsep]
    \item Space sciences
    \item Physics
    \item Chemistry
    \item Earth sciences (geology, oceanography, meteorology)
    \item Life science (ecology, zoology, botany)
    \item Life science (biology, genetics)
    \item Brain and neurosciences
    \item Medicine
    \item Engineering (civil, mechanical, chemical)
    \item Computer science and electrical engineering
    \item Economics
    \item Humanities and social sciences
    \item Tech support, lab tech, or support programmer
    \item Administration
    \item Other:
  \end{itemize}

\item
  In three sentences or less, please describe your current field of
  work or your research question.

\item
  What OS will you use on the laptop you bring to the workshop?
  \begin{itemize}[noitemsep]
    \item Linux
    \item Apple OS X
    \item Windows
    \item I do not know what operating system I use.
  \end{itemize}

\item
  With which programming languages, if any, could you write a program
  from scratch which imports some data and calculates mean and
  standard deviation of that data?
  \begin{itemize}[noitemsep]
    \item C
    \item C++
    \item Perl
    \item MATLAB
    \item Python
    \item R
    \item Java
    \item Other:
  \end{itemize}

\item
  What best describes how often you currently program?
  \begin{itemize}[noitemsep]
    \item I have never programmed.
    \item I program less than one a year.
    \item I program several times a year.
    \item I program once a month.
    \item I program once a week or more.
  \end{itemize}

\item
  What best describes the complexity of your programming? (Choose all
  that apply.)
  \begin{itemize}[noitemsep]
    \item I have never programmed.
    \item I write scripts to analyze data.
    \item I write tools to use and that others can use.
    \item I am part of a team which develops software.
  \end{itemize}

  \item
    A tab-delimited file has two columns showing the date and the
    highest temperature on that day. Write a program to produce a
    graph showing the average highest temperature for each month.
    \begin{itemize}[noitemsep]
    \item Could not complete.
    \item Could complete with documentation or search engine help.
    \item Could complete with little or no documentation or search engine help.
    \end{itemize}

  \item
    How familiar are you with Git version control?
    \begin{itemize}[noitemsep]
    \item Not familiar with Git.
    \item Only familiar with the name.
    \item Familiar with Git but have never used it.
    \item Familiar with Git because I have used or am using it.
    \end{itemize}

  \item
    Consider this task: given the URL for a project on GitHub, check
    out a working copy of that project, add a file called notes.txt,
    and commit the change.
    \begin{itemize}[noitemsep]
    \item Could not complete.
    \item Could complete with documentation or search engine help.
    \item Could complete with little or no documentation or search engine help.
    \end{itemize}

  \item
    How familiar are you with unit testing and code coverage?
    \begin{itemize}[noitemsep]
    \item Not familiar with unit testing or code coverage.
    \item Only familiar with the terms.
    \item Familiar with unit testing or code coverage but have never used it.
    \item Familiar with unit testing or code coverage because I have used or am using them.
    \end{itemize}

  \item
    Consider this task: given a 200-line function to test, write half
    a dozen tests using a unit testing framework and use code coverage
    to check that they exercise every line of the function.
    \begin{itemize}[noitemsep]
    \item Could not complete.
    \item Could complete with documentation or search engine help.
    \item Could complete with little or no documentation or search engine help.
    \end{itemize}

  \item
    How familiar are you with SQL?
    \begin{itemize}[noitemsep]
    \item Not familiar with SQL.
    \item Only familiar with the name.
    \item Familiar with SQL but have never used it.
    \item Familiar with SQL because I have used or am using them.
    \end{itemize}

  \item
    Consider this task: a database has two tables: Scientist and
    Lab. Scientist's columns are the scientist's user ID, name, and
    email address; Lab's columns are lab IDs, lab names, and scientist
    IDs. Write an SQL statement that outputs the number of scientists
    in each lab.
    \begin{itemize}[noitemsep]
    \item Could not complete.
    \item Could complete with documentation or search engine help.
    \item Could complete with little or no documentation or search engine help.
    \end{itemize}

  \item
    How familiar do you think you are with the command line?
    \begin{itemize}[noitemsep]
    \item Not familiar with the command line.
    \item Only familiar with the term.
    \item Familiar with the command line but have never used it.
    \item Familiar with the command line because I have or am using it.
    \end{itemize}

  \item
    How would you solve this problem: A directory contains 1000 text
    files. Create a list of all files that contain the word
    ``Drosophila'' and save the result to a file called results.txt.
    \begin{itemize}[noitemsep]
    \item Could not create this list.
    \item Would create this list using ``Find in Files'' and ``copy and paste''.
    \item Would create this list using basic command line programs.
    \item Would create this list using a pipeline of command line programs.
    \end{itemize}

  \end{itemize}

\end{document}